\documentclass[reprint,aps,pre]{revtex4-2}
\usepackage{graphicx,bm}
\usepackage{tipa,color}

\begin{document}

\title{Weighted average temperature as the effective temperature of a system in contact with two thermal baths}
\author{Z. C. Tu}\email[Email: ]{tuzc@bnu.edu.cn}
\affiliation{School of Physics and Astronomy, Beijing Normal University, Beijing 100875, China}
\affiliation{Key Laboratory of Multiscale Spin Physics (Ministry of Education), Beijing Normal University, Beijing 100875, China}

\date{\today}

\begin{abstract}
We investigate the effective temperature of a harmonic chain whose two ends are coupled to two baths at different temperatures. We propose to take the weighted average temperature as the effective temperature of the system. The weight factors are related to the couplings between the system and two baths as well as the asymmetry of interactions between oscillators. We revisit the thermodynamics of nonequilibrium steady states based on the weighted average temperature. It is found that the fundamental thermodynamic relations in nonequilibrium steady states possess similar concise forms as those in equilibrium thermodynamics, provided that we replace the temperature in equilibrium with the weighted average temperature in steady states. We also illustrate the procedure to explicitly calculate the effective temperatures via three examples. 
\end{abstract}
\maketitle

\section{Introduction}
Thermodynamics was regarded as a universal theory by Einstein~\cite{Klein1967}: ``It is the only physical theory of universal content, which I am convinced, that within the framework of applicability of its basic concepts will never be overthrown." The whole theoretical system of thermodynamics is on the basis of four laws of thermodynamics. Temperature is a core concept for the zeroth law of  thermodynamics, which characterizes an equilibrium state. However, this concept cannot be directly extended to general nonequilibrium states since the transitivity of thermal equilibrium is broken out of equilibrium. It is for this reason that has given rise to extensive discussions on the issue of temperature in nonequilibrium systems. McLennan stated: ``Nonequilibrium temperature is introduced for theoretical convenience rather than to take advantage of a basic principle"~\cite{McLennan1989}. Different terminologies of nonequilibrium temperatures such as kinetic temperature, local temperature, effective temperature have been proposed in various situations involving from passive to active, classical to quantum systems~\cite{Jou2003,Cugliandolo1997,Cugliandolo2011,Puglisi2017,DiVentra2019,Fielding2002,Ben-Isaac2011,Jabbari-Farouji2008,HayashiSasa2004,Nagel2004,Haxton2007,XuOHern2005,Abate2008,Ciliberto2009,Caso2010,Joly2011,Seifert2012,ZhangSzamel2011,Dieterich-NP971,Langer-PRE064139}.
Analogy to kinetic temperature for thermal systems, ``granular temperature" was also proposed for athermal systems such as granular materials, which is defined as the mean-square value of the random velocities of granular particles~\cite{Ogawa78}. Although the application of granular temperature in granular gases and fluids has proven to be great successful~\cite{Campbell-ARFM57,Goldhirsch-ARFM267,Goldhirsch-PT130}, the concept of granular temperature still confronts much debate~\cite{IppolitoPRE2017,Grossman-PRE4200,Losert-Chaos682,Olafsen-PRE2468,Baxter-GM135,Losert-PRL1428,RouyerPRL3676,Hayakawa-PRL2833,Hayakawa-PRL098001,Menon-PRL198301,WangPRE031301,Makse-Nat614,Shokef-PRE051111,Wangyj-Softmater5398,To-PRE062111} due to the lack of energy equipartition in granular systems.

Here we will confine our treatment to a less controversial situation of steady-state thermal systems in contact with two thermal baths at different temperatures, where the effective temperature is believed to be well-defined. There are two main viewpoints on effective temperatures of these systems in literature: (i) the mean temperature; (ii) the weighted average temperature.
The mean temperature is a natural definition of effective temperature for a system in contact with two baths at different temperatures. Parrondo and Espa\~{n}ol considered an axle with vanes at both ends symmetrically coupled to two baths, and found that the steady state can be described as the Boltzmann canonical distribution with an effective temperature being the arithmetical mean of temperatures of two baths~\cite{Parrondo1996}. In recent work by Wu and Wang~\cite{Wu-Wang2022}, the nonequilibrium equation of state of a harmonic chain coupled to two baths was established, which is divided into two parts: One has the form of the equilibrium equation of state with the equilibrium temperature replaced by the mean temperature of two baths; the other depends on the temperature difference between two baths. For a finite-size quantum system connected to two thermal and particle reservoirs, the nonequilibrium density matrix was derived by Ness, which is given by a generalized Gibbs-like ensemble with an effective reciprocal temperature being the mean of reciprocal temperatures of two baths~\cite{Ness2014}.

If the system is asymmetrically coupled to two baths, the weighted average temperature may be regarded as a more reasonable definition of effective temperature. Van den Broeck \textit{et al.}  investigated the heat transfer by a shared piston simultaneously in contact with two baths at different temperatures~\cite{VandenBroeck2001}. They found that the steady state is described by a canonical distribution with an effective temperature being the weighted average of temperatures of two baths. The weight factors are related to frictional coefficients of the piston in both baths. The idea of weighted average temperature holds also for underdamped Langevin systems in contact with multiple reservoirs~\cite{Esposito2016,Park2018} and quantum heat transfers at the nanoscale~\cite{CaoJ2019}. Sheng and the author in the present paper also suggested the weighted average reciprocal temperature as the  inverse effective temperature of finite-time heat engines, the devices outputting work and absorbing heat simultaneously~\cite{ShengTuPRE14}. The weighted average temperatures mentioned above are introduced for simple systems with few degrees of freedom. We may ask: whether can we give a unified definition and calculation procedure of effective temperature in steady state via weighted average temperature for a system with multiple degrees of freedom? Can the fundamental relations in equilibrium thermodynamics be extended to nonequilibrium steady states? Do the extended relations possess the same forms as those in equilibrium thermodynamics?

Aiming at the questions mentioned above, we will develop the idea of weighted average temperature based on a model of harmonic chain coupled to two thermal baths. 
This model has been widely used in discussing heat conduction of low-dimensional systems. Rieder \textit{et al.} discussed a stationary state of a homogeneous harmonic chain coupled symmetrically to two baths at different temperatures, and found that the kinetic temperature of the chain is almost the mean temperature of two baths, and that the heat flux is proportional to the global temperature difference of two baths but not the local temperature gradient in the chain~\cite{Rieder-Lebowitz1967}. Matsuda and Ishii considered an isotopically disordered harmonic chain with infinite length, and found that the heat flux depends on the square root of the number of oscillators in the chain~\cite{Matsuda-IshiiPTPS1970}. Then the heat transport in low-dimensional systems becomes one of the most exciting topics in the field of statistical physics, and intense researches ~\cite{Lebowitz19711701,SpohnCMP1977,KipnisJSP1982,Casati198651,LepriPRL1896,DharPRL1999480,ProsenPRL2000,LiZhaoHu2001,DharPRL2001,SavinPRL2002,ZhouXin061202,NarayanPRL2002,LiBW-Wangj2003,Casati2005015120,ZhaoHongPRL2006,Roy2008535,RenJPRE2010,KannanDharPRE2012,ZhongZhaoPRE060102,HanggiPRL2014,Xiong2017042109,Sato20202012111,Weiderpass102125401,Zhao2020186401,Onorato2020034110,WangleiPRE2024} are focused on the universality class of the divergence of thermal conductivity, the profile of local temperature in the chain, the role of disorder and localization in low-dimensional lattices. The reader may refer to four reviews~\cite{Lepri-Livi-Politi2003,Dhar-AdvPhys2008,LiBaowenRMP2018,Livi2023127779} to gain a comprehensive survey on heat transport in low-dimensional systems. Here we will focus on another issue: Can we define the effective temperature for the whole system at steady state? This issue can be discussed independently of whether or not the thermal conductivity diverges, the local equilibrium holds, and normal modes are localized.

We will try to give a unified definition of effective temperature and then revisit thermodynamics of nonequilbrium steady states via the weighted average temperature.
The rest of this paper is organized as follows. In Sec.~\ref{sec-Modelsys}, we present a minimal model which consists of a harmonic chain with its two ends coupled to two thermal baths at different temperatures. In Sec.~\ref{sec-WATemp}, we introduce the idea of weighted average temperature according to a decomposition of covariance matrix. In Sec.~\ref{sec-steadythermod}, we revisit thermodynamics of nonequilibrium steady states by taking the weighted average temperature as the effective temperature. In Sec.~\ref{sec-Casestudy}, we discuss three examples and explicitly calculate the corresponding effective temperatures. The last section contains a brief summary and discussion.

\section{Model system: a harmonic chain coupled to two thermal baths\label{sec-Modelsys}}
\begin{figure}[htp!]
\includegraphics[width=8cm]{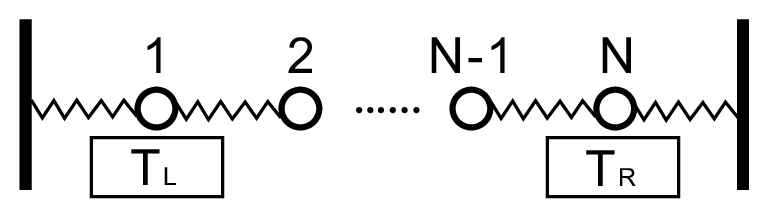}
\caption{A harmonic chain with its two ends coupled to two thermal baths at different temperatures $T_L$ and $T_R$.}\label{fig-harmchain}
\end{figure}

The model system that we considered is a harmonic chain coupled to two baths as shown in Fig.~\ref{fig-harmchain}. The chain consists of $N$ particles and $N+1$ elastic springs. Particle 1 is coupled to a bath at temperature $T_L$, and particle $N$ is in contact with a bath at temperature $T_R$.
The equations of motion of particle $i$ ($i=1,2,3,\cdots,N$) may be expressed as the Langevin dynamics:
\begin{eqnarray}\frac{d x_i}{d t}&=&{p_i},\label{eq-dxdt}\\
\frac{d p_i}{d t}&=&-\bar{k}_{i-1}x_{i-1}-k_{i}x_i-\bar{k}_{i}x_{i+1}\nonumber \\
&+&[-\gamma_L p_1+\xi_L(t)]\delta_{i1}+[-\gamma_R p_N+\xi_R(t)]\delta_{iN},\label{eq-dpdt}
\end{eqnarray}
where $x_i$ and $p_i$ represent the coordinate and momentum of particle $i$, respectively. In Eq.~(\ref{eq-dxdt}), we have assumed that all particles are of the same mass which is set to be unity for simplicity. In Eq.~(\ref{eq-dpdt}), $k_i$ and $\bar{k}_i$ are related to the elastic constants of the springs. We have imposed $\bar{k}_{0}=\bar{k}_{N}=0$. $\delta_{i1}$ and $\delta_{iN}$ are the Kronecker delta notations. $\gamma_L$ and $\gamma_R$ represent frictional coefficients of particles 1 and $N$, respectively. $\xi_L(t)$ and $\xi_R(t)$ are white noises due to the thermal baths, which satisfy $\langle \xi_{\alpha}(t)\rangle=0$ and $\langle \xi_{\alpha}(t)\xi_{\alpha}(t')\rangle=\gamma_\alpha k_BT_\alpha\delta(t-t')$ with $\alpha=L$ or $R$. $k_B$ is the Boltzmann constant which is set to be unity in the following discussions.

Introduce position vector $\mathbf{x}=[x_1,x_2,\cdots,x_N]^\mathrm{T}$, momentum vector $\mathbf{p}=[p_1,p_2,\cdots,p_N]^\mathrm{T}$, phase vector $\mathbf{z}=[x_1,x_2,\cdots,x_N,p_1,p_2,\cdots,p_N]^\mathrm{T}$, and noise vector $\bm{\xi}=[0,0,\cdots,0,\xi_1,0\cdots,0,\xi_N]^\mathrm{T}$.  In the whole paper, the superindex $\mathrm{T}$ represents the transpose of a matrix.
Introduce $N\times N$-order elastic matrix
\begin{equation}\label{eq-elasticmatrix}
\mathbf{K}=\left[\begin{array}{cccccc}k_1 & \bar{k}_1 & 0 & 0 & \cdots & 0 \\ \bar{k}_1 & k_2 & \bar{k}_2 & 0 & \cdots & 0 \\ 0 & \bar{k}_2 & k_3 & \bar{k}_3 & \cdots & 0 \\ \vdots & \ddots & \ddots & \ddots & \ddots & \vdots \\ 0 & \cdots & 0 & \bar{k}_{N-2} & k_{N-1} & \bar{k}_{N-1} \\ 0 & 0 & \cdots & 0 & \bar{k}_{N-1} & k_N\end{array}\right],
\end{equation}
and frictional matrix
\begin{equation}\label{eq-frictmatrix}
\bm{\Gamma}=\left[\begin{array}{ccccc}\gamma_L & 0 & \cdots &0& 0 \\ 0 & 0 & \cdots & 0&0 \\ \vdots & \vdots &\ddots & \vdots & \vdots \\ 0 & 0&\cdots & 0 & 0 \\ 0 & 0&\cdots & 0 & \gamma_R\end{array}\right].
\end{equation}
Then the equations of motion of the system may be transformed into a matrix form:
\begin{equation}
\frac{d\mathbf{z}}{dt}=-\mathbf{A}\mathbf{z}+\bm{\xi}\label{eq-Langven-matrx},
\end{equation}
where coefficient matrix
\begin{equation}
\mathbf{A}=\left[\begin{array}{cc}\mathbf{0} & -\mathbf{I} \\ \mathbf{K} & \bm{\Gamma} \end{array}\right]\end{equation} with $\mathbf{0}$ and $\mathbf{I}$ being $N\times N$-order zero matrix and identity matrix, respectively.

Introduce steady-state covariance matrices $\bm{\sigma}_{xx}=\langle\mathbf{x}\mathbf{x}^\mathrm{T}\rangle_{t\rightarrow \infty}$, $\bm{\sigma}_{xp}=\langle\mathbf{x}\mathbf{p}^\mathrm{T}\rangle_{t\rightarrow \infty}$, $\bm{\sigma}_{px}=\langle\mathbf{p}\mathbf{x}^\mathrm{T}\rangle_{t\rightarrow \infty}$, $\bm{\sigma}_{pp}=\langle\mathbf{p}\mathbf{p}^\mathrm{T}\rangle_{t\rightarrow \infty}$, and
$\bm{\sigma}=\langle \mathbf{z}\mathbf{z}^\mathrm{T} \rangle_{t\rightarrow \infty}=\left[\begin{array}{cc}\bm{\sigma}_{xx} & \bm{\sigma}_{xp} \\ \bm{\sigma}_{px} & \bm{\sigma}_{pp}\end{array}\right]$.
Covariance matrix $\bm{\sigma}$ is symmetrical, which implies $\bm{\sigma}_{xx}^\mathrm{T}=\bm{\sigma}_{xx}$, $\bm{\sigma}_{pp}^\mathrm{T}=\bm{\sigma}_{pp}$ and $\bm{\sigma}_{xp}^\mathrm{T}=\bm{\sigma}_{px}$.

Since Eq.~(\ref{eq-Langven-matrx}) describes an Ornstein-Uhlenbeck process, the covariance matrix satisfies the continuous-time Lyapunov equation~\cite{Cugliandolo2011,Wu-Wang2022,Rieder-Lebowitz1967,Gardinerbook,Riskenbook}:
\begin{equation}\mathbf{A}\bm{\sigma}+\bm{\sigma}\mathbf{A}^\mathrm{T}=2\mathbf{D}_\mathrm{t},\label{eq-Lyapunov}
\end{equation}
where $\mathbf{D}_\mathrm{t}=\left[\begin{array}{cc}\mathbf{0} & \mathbf{0}\\ \mathbf{0} & \mathbf{D}\end{array}\right]$. $\mathbf{D}$ is $N\times N$-order diffusion matrix which is explicitly expressed as
\begin{equation}
\mathbf{D}=\left[\begin{array}{ccccc}\gamma_L T_L & 0 & \cdots &0& 0 \\ 0 & 0 & \cdots & 0&0 \\ \vdots & \vdots &\ddots & \vdots & \vdots \\ 0 & 0&\cdots & 0 & 0 \\ 0 & 0&\cdots & 0 & \gamma_RT_R\end{array}\right]\label{eq-diffusionmatrix}.
\end{equation}

\section{Weighted average temperature\label{sec-WATemp}}
In this section, we will introduce a parameter $T_e$ with dimension of temperature via the weighted average of temperatures $T_L$ and $T_R$. This parameter can be regarded as a candidate for the effective temperature of a system coupled to two baths at different temperatures. 

If the two baths have the same temperature $T$, i.e., $T_L=T_R=T$, the system will reach an equilibrium state. The theorem of energy equipartition leads to $\bm{\sigma}_{xx}\mathbf{K}=\bm{\sigma}_{pp}=T \mathbf{I}$ where the Boltzmann constant has been set to be unity. With the consideration of this point, we introduce $\tilde{\mathbf{K}}=\left[\begin{array}{ll}\mathbf{K} & \mathbf{0} \\ \mathbf{0} & \mathbf{I}\end{array}\right]$ and define revised covariance matrix
\begin{equation}\label{revcovMatrix}
\tilde{\bm\sigma}={\bm\sigma} \tilde{\mathbf{K}}=\left[\begin{array}{ll}\bm\sigma_{xx} \mathbf{K} & \bm\sigma_{xp} \\ \bm\sigma_{px} \mathbf{K} & \bm\sigma_{p p}\end{array}\right]
\end{equation}
such that equilibrium covariance matrix $\tilde{\bm\sigma}_{eq}=T \mathbf{I}_{2N}$ with $\mathbf{I}_{2N}$ being $2N\times 2N$-order identity matrix. This observation inspires us to make an ansatz that the revised covariance matrix for nonequlibrium steady states may be decomposed into a diagonal matrix plus a residual matrix. More specifically, we introduce a parameter $T_e$ with dimension of temperature for nonequlibrium steady states such that
\begin{equation}\label{eq-effecttemp1}
\tilde{\bm{\sigma}}=T_{e}\mathbf{I}_{2N}+\tilde{\bm{\sigma}}^r,
\end{equation}
where the residual matrix is traceless:
\begin{equation}\label{eq-traceless}
\mathrm{Tr}\tilde{\bm{\sigma}}^r=0.
\end{equation}
Obviously, the above two equations hold for equilibrium states since $T_e=T$ and $\tilde{\bm{\sigma}}^r=\mathbf{0}$ at equilibrium states.

Introducing 
\begin{equation}\label{eq-tildeAmatrx}
\tilde{\mathbf{A}}=\tilde{\mathbf{K}} \mathbf{A}=\left[\begin{array}{cc}\mathbf{0} & -\mathbf{K} \\ \mathbf{K} &\bm{\Gamma}\end{array}\right]
\end{equation}
 and considering Eqs.~(\ref{eq-Lyapunov}) and (\ref{eq-effecttemp1}), we find that the residual matrix should satisfy the following equation:
\begin{equation}
\tilde\mathbf{A}\tilde{\bm{\sigma}}^r+\tilde{\bm{\sigma}}^{r\mathrm{T}}\tilde\mathbf{A}^\mathrm{T}=2\left[\begin{array}{ll}\mathbf{0} & \mathbf{0} \\ \mathbf{0} & \mathbf{D}-T_{e}\bm{\Gamma}\end{array}\right],\label{eq-resmatrix}
\end{equation}
where $\bm{\Gamma}$ and $\mathbf{D}$ are given by Eqs.~(\ref{eq-frictmatrix}) and (\ref{eq-diffusionmatrix}), respectively. The detailed derivation of (\ref{eq-resmatrix}) is demonstrated in Appendix~\ref{sec-ap-12to13}.
The above equation is linear, therefore its solution can be expressed as a linear combination of two bases
$\tilde{\bm{\sigma}}_L$ and $\tilde{\bm{\sigma}}_R$ which satisfy
\begin{equation}\label{eq-sigmaLR}
\tilde\mathbf{A}\tilde{\bm{\sigma}}_\alpha+\tilde{\bm{\sigma}}_\alpha^\mathrm{T}\tilde\mathbf{A}^\mathrm{T}=2\left[\begin{array}{ll}\mathbf{0} & \mathbf{0} \\ \mathbf{0} & \mathbf{E}_\alpha\end{array}\right],~(\alpha=L~\mathrm{or}~R),
\end{equation}
where $\mathbf{E}_L$ and $\mathbf{E}_R$ are two $N\times N$-order matrix units. Their explicit forms are as follows:
\begin{equation}
\mathbf{E}_L=\left[\begin{array}{cccc}1 & 0 & \cdots &0 \\ 0 & 0 & \cdots & 0\\ \vdots & \vdots &\ddots & \vdots  \\ 0 & 0&\cdots & 0 \end{array}\right],~\mathbf{E}_R=\left[\begin{array}{ccccc}0 & \cdots & 0&0 \\ \vdots &\ddots & \vdots & \vdots \\  0&\cdots & 0 & 0 \\ 0&\cdots & 0 & 1\end{array}\right].
\end{equation}

Once we obtain $\tilde{\bm{\sigma}}_L$ and $\tilde{\bm{\sigma}}_R$ from Eq.~(\ref{eq-sigmaLR}), the residual matrix may be expressed as
\begin{equation}\label{eq-linearcombin}
\tilde{\bm{\sigma}}^r=\gamma_L(T_L-T_{e})\tilde{\bm{\sigma}}_L+\gamma_R(T_R-T_{e})\tilde{\bm{\sigma}}_R.
\end{equation}
By taking the trace in the above equation and considering Eq.~(\ref{eq-traceless}), we obtain
\begin{equation}\label{eq-effectivetemp}
T_{e}=C_LT_L+C_RT_R,
\end{equation}
where coefficients $C_L$ and $C_R$ satisfy
\begin{equation}\label{eq-weightfactor}
C_\alpha\equiv\frac{\gamma_\alpha\mathrm{Tr}\tilde{\bm{\sigma}}_\alpha}{\gamma_L\mathrm{Tr}\tilde{\bm{\sigma}}_L+\gamma_R\mathrm{Tr}\tilde{\bm{\sigma}}_R},~(\alpha=L~\mathrm{or}~R).
\end{equation}
Since $C_L+C_R=1$, temperature $T_{e}$ is the weighted average of temperatures $T_L$ and $T_R$. Thus $T_e$ is named weighted average temperature in this paper.
Substituting Eqs.~(\ref{eq-effectivetemp}) and (\ref{eq-weightfactor}) into Eq.~(\ref{eq-linearcombin}), we obtain the residual matrix
\begin{equation}\label{eq-sigma-r}
\tilde{\bm{\sigma}}^r=\gamma_L\gamma_R \Delta T\frac{\tilde{\bm{\sigma}}_L\mathrm{Tr}\tilde{\bm{\sigma}}_R-\tilde{\bm{\sigma}}_R\mathrm{Tr}\tilde{\bm{\sigma}}_L}{\gamma_L\mathrm{Tr}\tilde{\bm{\sigma}}_L+\gamma_R\mathrm{Tr}\tilde{\bm{\sigma}}_R},
\end{equation}
which is proportional to the temperature difference $\Delta T\equiv T_L-T_R$.

We emphasize that the weighted average reciprocal temperature was also introduced as the inverse effective temperature for finite-time heat engines~\cite{ShengTuPRE14}. That is, $1/T_e \equiv s_L/T_L+s_R/T_R$ with $s_L+s_R=1$. This definition is in fact equivalent to the weighted average temperature in the present work. Compared with Eq.~(\ref{eq-effectivetemp}), we can derive a duality relation between $(C_L,C_R)$ and $(s_L,s_R)$ which may be expressed as $C_L={s_L T_R}/(s_L T_R+s_R T_L)$ and $C_R={s_R T_L}/(s_L T_R+s_R T_L)$.

\section{Thermodynamics of Nonequilibrium Steady States\label{sec-steadythermod}}
In this section, we will revisit thermodynamics of nonequilibrium steady state of the harmonic chain coupled to two thermal baths
by using the weighted average temperature. Several fundamental thermodynamic expressions that we obtained possess similar concise forms as those in equilibrium thermodynamics. In this sense, we propose to take the weighted average temperature as the effective temperature of the system.

\subsection{Steady-state distribution\label{subsec-distrib}}
The stochastic dynamics (\ref{eq-Langven-matrx}) describes an Ornstein-Uhlenbeck process. The steady-state distribution is Gaussian
distribution~\cite{Rieder-Lebowitz1967,Riskenbook}:
\begin{equation}\label{eq-steadydistrib1}
P(\mathbf{z})=\frac{1}{\mathcal{Z}} \exp \left\{-\frac{1}{2} \mathbf{z}^\mathrm{T} \bm{\sigma}^{-1} \mathbf{z}\right\}
\end{equation}
where the partition function is
\begin{equation}\label{eq-partitionfunct1}
\mathcal{Z}=\sqrt{\mathrm{det}\left(2 \pi \bm{\sigma}\right)}.
\end{equation}
Note that we have omitted the term related to the Planck constant in this work.

Considering Eqs.~(\ref{revcovMatrix}), (\ref{eq-effecttemp1}), and (\ref{eq-sigma-r}), we may rewrite the steady-state distribution
with inverse weighted average temperature $\beta_e=1/T_e$, which reads
\begin{equation}\label{eq-steadydistrib}
P(\mathbf{z})=\frac{1}{\mathcal{Z}} \exp \left\{-\beta_e (H+\Delta H)\right\},
\end{equation}
where the Hamiltonian 
\begin{equation}H=\frac{1}{2}\mathbf{p}^\mathrm{T}\mathbf{p}+\frac{1}{2}\mathbf{x}^\mathrm{T}\mathbf{Kx}=\frac{1}{2}\mathbf{z}^\mathrm{T}\tilde\mathbf{K}\mathbf{z}.
\end{equation}
The additional Hamiltonian is
\begin{eqnarray}
\Delta H&=&\frac{1}{2}\mathbf{z}^\mathrm{T} \tilde{\mathbf{K}}\left[\left(\mathbf{I}+\beta_{e}\tilde{\bm{\sigma}}^r\right)^{-1}-\mathbf{I}\right] \mathbf{z}\nonumber \\
&\approx&\frac{\gamma_L\gamma_R\Delta T}{2T_{e}}\mathbf{z}^\mathrm{T} \tilde{\mathbf{K}}\left[\frac{\tilde{\bm{\sigma}}_L\mathrm{Tr}\tilde{\bm{\sigma}}_R-\tilde{\bm{\sigma}}_R\mathrm{Tr}\tilde{\bm{\sigma}}_L}{\gamma_L\mathrm{Tr}\tilde{\bm{\sigma}}_L+\gamma_R\mathrm{Tr}\tilde{\bm{\sigma}}_R} \right] \mathbf{z}.
\end{eqnarray}
The approximation in the second line of the above equation holds for small temperature difference. The steady-state distribution function (\ref{eq-steadydistrib}) and the linear dependence of $\Delta H$ on $\Delta T$ are consistent to those obtained in Refs.~\cite{Ness2014} and \cite{CaoJ2019}.
In addition, considering Eqs.~(\ref{eq-effecttemp1}) and (\ref{eq-traceless}), we can further derive the expression of partition function for small temperature difference, which reads
\begin{equation}\label{eq-partitionfunct2}
\mathcal{Z}=\left(2 \pi T_e\right)^N+\mathcal{O}(\Delta T^2/T_e^2).
\end{equation}
That is, there is no explicitly linear term of $\Delta T/T_e$ in the expression of partition function.

\subsection{Internal energy\label{subsec-intenergy}}
The steady-state internal energy is defined as an average of the Hamiltonian:
\begin{equation}\label{eq-intenergy}
\mathcal{E}\equiv \langle H\rangle =\frac{1}{2}\langle\mathbf{p}^\mathrm{T}\mathbf{p}\rangle +\frac{1}{2}\langle\mathbf{x}^\mathrm{T}\mathbf{Kx}\rangle.
\end{equation}

Since $\mathbf{p}=[p_1,p_2,\cdots,p_N]^\mathrm{T}$ is a column vector, $\mathbf{p}^\mathrm{T}\mathbf{p}=\sum_{i=1}^N p_i^2$ is a pure number, while $\mathbf{p}\mathbf{p}^\mathrm{T}$ is a matrix with an element $p_ip_j$ at row $i$ and column $j$. Thus we arrive at $\langle\mathbf{p}^\mathrm{T}\mathbf{p}\rangle=\mathrm{Tr}\langle\mathbf{p}\mathbf{p}^\mathrm{T}\rangle=\mathrm{Tr}\bm{\sigma}_{pp}$. Similarly, $\mathbf{x}=[x_1,x_2,\cdots,x_N]^\mathrm{T}$ is a column vector, $\mathbf{x}^\mathrm{T}\mathbf{Kx}=\sum_{i=1}^N\sum_{j=1}^N  K_{ij}x_ix_j$ is also a pure number. Its average $\langle\mathbf{x}^\mathrm{T}\mathbf{Kx}\rangle=\sum_{i=1}^N\sum_{j=1}^N  K_{ij}\langle x_ix_j\rangle=\mathrm{Tr}(\bm{\sigma}_{xx}\mathbf{K})$. Substituting these relations into Eq.~(\ref{eq-intenergy}), we obtain the internal energy
$\mathcal{E}\equiv \langle H\rangle =[\mathrm{Tr}(\bm{\sigma}_{xx}\mathbf{K})+\mathrm{Tr}\bm{\sigma}_{pp}]/2=\mathrm{Tr}\tilde{\bm{\sigma}}/2$. With the consideration of Eqs.~(\ref{eq-effecttemp1}) and (\ref{eq-traceless}), we arrive at the internal energy
\begin{equation}\label{eq-intenergy2}
\mathcal{E}=N T_e.
\end{equation}
This concise relation for nonequilibrium steady states keeps the same form in equilibrium states provided that we replace the temperature in equilibrium with the weighted average temperature $T_e$. In addition, we can further verify $T_e=\mathrm{Tr}\bm{\sigma}_{pp}/2$ using the conclusion in Appendix~\ref{equaltraceBG}. That is, $T_e$ reflects the average kinetic energy of the system. Therefore, we deduce that $T_e$ has the meaning of effective temperature. In this sense, our proposal is consistent with the ideas of kinetic temperature or granular temperature in literature~\cite{Jou2003,Puglisi2017,Abate2008,Seifert2012,ZhangSzamel2011,Losert-Chaos682,Menon-PRL198301,To-PRE062111,Shokef-PRE051111,IppolitoPRE2017,Grossman-PRE4200,Baxter-GM135,Hayakawa-PRL098001,Ogawa78,Campbell-ARFM57,Goldhirsch-ARFM267,Goldhirsch-PT130}.

\subsection{Entropy and free energy}
 The steady-state entropy is defined as
 $\mathcal{S}=-\int d\mathbf{x}d\mathbf{p} P\ln P=-\langle \ln P\rangle$. Substituting the steady-state distribution (\ref{eq-steadydistrib1}) into the above equation, we derive the entropy
 $\mathcal{S}=\langle   \mathbf{z}^\mathrm{T} \bm{\sigma}^{-1} \mathbf{z} \rangle /2+\ln\mathcal{Z}$. It is not hard to prove $\langle   \mathbf{z}^\mathrm{T} \bm{\sigma}^{-1} \mathbf{z} \rangle = \sum_{i=1}^N\sum_{j=1}^N [\sigma]^{-1}_{ij}\langle z_iz_j\rangle=\mathrm{Tr}(\bm\sigma^{-1}\bm\sigma)=2N$. Considering expression (\ref{eq-intenergy2}) of internal energy, we may derive 
 \begin{equation}\label{eq-entropy}
 \mathcal{S}=\frac{\mathcal{E}}{T_e} + \ln\mathcal{Z}.
\end{equation}

Similar to the equilibrium state, we may define the steady-state free energy $\mathcal{F}=\mathcal{E}-T_e \mathcal{S}$. Thus Eq.~(\ref{eq-entropy}) leads to
\begin{equation}\label{eq-freeenergy}
\mathcal{F}=-T_e \ln\mathcal{Z}.
\end{equation}
This concise relation for nonequilibrium steady states keeps the same form in equilibrium states provided that we replace the temperature in equilibrium with the weighted average temperature. This is another fact that supports us to take the weighted average temperature $T_e$ as the effective temperature of the system.

\subsection{Heat transfer and entropy production}
The rate of heat transfer from the baths at temperature $T_L$ and  $T_R$ to the chain may be defined as~\cite{Parrondo1996,Wu-Wang2022,VandenBroeck2001,Rieder-Lebowitz1967,Matsuda-IshiiPTPS1970}:
\begin{eqnarray}\label{eq-heattrans1}
\dot\mathcal{Q}_L=\gamma_L(T_L-\langle p_1^2 \rangle),\\
\dot\mathcal{Q}_R=\gamma_R(T_R-\langle p_N^2 \rangle),
\end{eqnarray}
respectively. It is not hard to verify $\dot\mathcal{Q}_L+\dot\mathcal{Q}_R=0$ since $\mathrm{Tr}(\bm{\Gamma\sigma}_{pp})=\mathrm{Tr}\mathbf{D}$, which
is consistent to energy conservation in steady states~\cite{Wu-Wang2022}.

According to Eqs.~(\ref{eq-effecttemp1}) and (\ref{eq-sigma-r}), we find that both $\langle p_1^2 \rangle$ and $\langle p_N^2 \rangle$ are different from $T_e$ in the linear order of $\Delta T$. They are explicitly expressed as $\langle p_1^2 \rangle=T_e+u_L\Delta T$ and $\langle p_N^2 \rangle=T_e-u_R\Delta T$ with two constants $u_L$ and $u_R$. Note that $u_L$ and $u_R$ are not independent of each other since $\dot\mathcal{Q}_L+\dot\mathcal{Q}_R=0$. According to Eq.~(\ref{eq-effectivetemp}), the rate of heat transfer may be further expressed as
\begin{eqnarray}\label{eq-heattrans2}
\dot\mathcal{Q}_L=\gamma_L(C_R-u_L)\Delta T,
\end{eqnarray}
which implies that the rate of heat transfer is proportional to the temperature difference $\Delta T\equiv T_L-T_R$. In addition, if $\Delta T>0$, then $\dot\mathcal{Q}_L>0$ which gives a constraint $C_R>u_L$.

The entropy production rate may be defined as
\begin{equation}\label{eq-entropyprodrate1}
\dot\mathcal{S}_p=-\left(\frac{\dot\mathcal{Q}_L}{T_L} +\frac{\dot\mathcal{Q}_R}{T_R} \right).
\end{equation}
Considering $\dot\mathcal{Q}_R=-\dot\mathcal{Q}_L$ and Eq.~(\ref{eq-heattrans2}), we obtain the entropy production rate:
\begin{equation}\label{eq-entropyprodrate2}
\dot\mathcal{S}_p=\frac{\gamma_L (C_R-u_L) \Delta T^2}{T_LT_R}.
\end{equation}
That is, the entropy production rate is proportional to the quadratic order term of temperature difference for given values of $T_L$ and $T_R$.

\section{Case study for calculation of effective temperatures\label{sec-Casestudy}}
In this section, we will illustrate the detailed procedure for explicit calculation of effective temperatures. The key step is to solve equation (\ref{eq-sigmaLR}). Here we adopt a method similar to the work by Rieder, Lebowitz and Lieb~\cite{Rieder-Lebowitz1967}.
The $2N\times 2N$-order matrix $\tilde{\bm{\sigma}}_\alpha$ may be expressed in block form: 
\begin{equation}\label{sigma-block}
\tilde{\bm{\sigma}}_\alpha=\left[\begin{array}{cc}\mathbf{B}_\alpha & \mathbf{J}_\alpha \\ \mathbf{F}_\alpha & \mathbf{G}_\alpha\end{array}\right],~(\alpha=L~\mathrm{or}~R),
\end{equation}
where $\mathbf{B}_\alpha$, $\mathbf{J}_\alpha$, $\mathbf{F}_\alpha$, and $\mathbf{G}_\alpha$ are four $N\times N$-order matrices.
Considering Eqs.~(\ref{revcovMatrix}) and (\ref{eq-sigmaLR}), we derive six equations as follows:
\begin{eqnarray}
&&\mathbf{G}^\mathrm{T}_\alpha=\mathbf{G}_\alpha,\label{eq-matrixG}\\
&&\mathbf{F}^\mathrm{T}_\alpha=\mathbf{KJ}_\alpha,\label{eq-matrixFJ}\\
&&(\mathbf{KB}_\alpha)^\mathrm{T}=\mathbf{KB}_\alpha,\label{eq-matrixKB}\\
&&(\mathbf{KF}_\alpha)^\mathrm{T}=-\mathbf{KF}_\alpha,\label{eq-matrixKF}\\
&&\mathbf{KB}_\alpha+\bm{\Gamma}\mathbf{F}_\alpha=\mathbf{G}_\alpha \mathbf{K},\label{eq-matrixKBG}\\
&&\mathbf{F}_\alpha+\mathbf{F}^\mathrm{T}_\alpha+\bm{\Gamma}\mathbf{G}_\alpha+\mathbf{G}_\alpha\bm{\Gamma}=2\mathbf{E}_\alpha,\label{eq-matrixF}
\end{eqnarray}
where $\alpha=L$ or $R$. Then the weight factors can be further expressed as
\begin{equation}\label{eq-weightfactor2aa}
C_L=\frac{\gamma_L(\mathrm{Tr}\mathbf{B}_L+\mathrm{Tr}\mathbf{G}_L)}{\gamma_L(\mathrm{Tr}\mathbf{B}_L+\mathrm{Tr}\mathbf{G}_L)+\gamma_R(\mathrm{Tr}\mathbf{B}_R+\mathrm{Tr}\mathbf{G}_R)}
\end{equation}
and
\begin{equation}\label{eq-weightfactor3aa}
C_R=\frac{\gamma_R(\mathrm{Tr}\mathbf{B}_R+\mathrm{Tr}\mathbf{G}_R)}{\gamma_L(\mathrm{Tr}\mathbf{B}_L+\mathrm{Tr}\mathbf{G}_L)+\gamma_R(\mathrm{Tr}\mathbf{B}_R+\mathrm{Tr}\mathbf{G}_R)}
\end{equation}
with the consideration of Eqs.~(\ref{eq-weightfactor}) and (\ref{sigma-block}).

Through tedious manipulations (as shown in Appendix~\ref{equaltraceBG}), we can prove $\mathrm{Tr}\mathbf{B}_L=\mathrm{Tr}\mathbf{G}_L$ and $\mathrm{Tr}\mathbf{B}_R=\mathrm{Tr}\mathbf{G}_R$. Then the above equations are further simplified as
\begin{equation}\label{eq-weightfactor2}
C_L=\frac{\gamma_L\mathrm{Tr}\mathbf{G}_L}{\gamma_L\mathrm{Tr}\mathbf{G}_L+\gamma_R\mathrm{Tr}\mathbf{G}_R}
\end{equation}
and
\begin{equation}\label{eq-weightfactor3}
C_R=\frac{\gamma_R\mathrm{Tr}\mathbf{G}_R}{\gamma_L\mathrm{Tr}\mathbf{G}_L+\gamma_R\mathrm{Tr}\mathbf{G}_R}.
\end{equation}

Now, we will discuss three examples: (A) Single harmonic oscillator simultaneously coupled to two baths; (B) Two harmonic oscillators coupled to two baths; (C) A chain of three harmonic oscillators with two ends coupled to two baths.

\subsection{Single harmonic oscillator simultaneously coupled to two baths}
\begin{figure}[htp!]
\includegraphics[width=8cm]{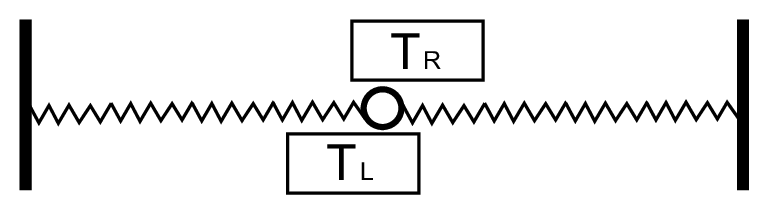}
\caption{Single harmonic oscillator simultaneously coupled to two baths at different temperatures $T_L$ and $T_R$.}\label{fig-Singleoscillator}
\end{figure}

Single harmonic oscillator simultaneously coupled to two baths as shown in Fig.~\ref{fig-Singleoscillator} is the special case of model system considered in Sec.~\ref{sec-Modelsys} with $N=1$. In this case, the elastic matrix, the frictional matrix and so on are degenerated into pure numbers. For example,
$\mathbf{K}=k$, $\bm{\Gamma}=\gamma_L+\gamma_R$, $\mathbf{E}_L=\mathbf{E}_R=1$. From Eqs.~(\ref{eq-matrixG})-(\ref{eq-matrixF}), we obtain
\begin{equation}
\mathbf{G}_{L}=\mathbf{G}_R=\frac{1}{\gamma_L+\gamma_R}.
\end{equation}
Substituting the above equation into Eqs.~(\ref{eq-weightfactor2}) and (\ref{eq-weightfactor3}), we arrive at
\begin{equation}
C_\alpha=\frac{\gamma_\alpha}{\gamma_L+\gamma_R},~(\alpha=L~\mathrm{or}~R).
\end{equation}
Substituting the above equation into (\ref{eq-effectivetemp}), we obtain the effective temperature
\begin{equation}
T_{e}=C_LT_L+C_RT_R=\frac{\gamma_LT_L+\gamma_RT_R}{\gamma_L+\gamma_R}.
\end{equation}
The above equation agrees with the result obtained by Van den Broeck \textit{et al.} in Ref.~\cite{VandenBroeck2001}. In particular,  $T_{e}=\bar{T}\equiv (T_L+T_R)/2$ if $\gamma_L=\gamma_R$, which degenerates into the result obtained by Parrondo and Espa\~{n}ol~\cite{Parrondo1996}.

\subsection{Two harmonic oscillators coupled to two baths\label{subsec-twooscillators}}
\begin{figure}[htp!]
\includegraphics[width=8cm]{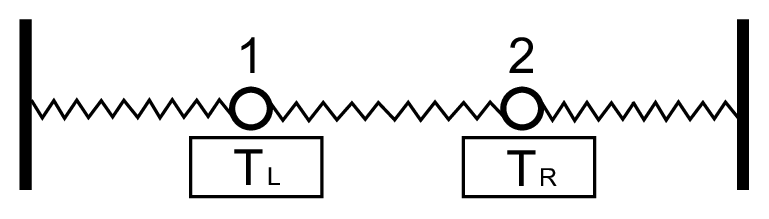}
\caption{Two harmonic oscillators coupled to two baths at different temperatures $T_L$ and $T_R$.}\label{fig-twooscillator}
\end{figure}

Two harmonic oscillators respectively coupled to two baths as shown in Fig.~\ref{fig-twooscillator} is the special case of model system considered in Sec.~\ref{sec-Modelsys} with $N=2$. In this case, the elastic matrix and the frictional matrix are assumed to be $\mathbf{K}=\left[\begin{array}{cc}k_1 & \bar{k}\\ \bar{k} & k_2 \end{array}\right]$ and $\bm\Gamma=\left[\begin{array}{cc}\gamma_L & 0 \\ 0 &  \gamma_R\end{array}\right]$.

Taking $\mathbf{E}_L=\left[\begin{array}{cc}1 & 0 \\ 0 & 0\end{array}\right]$, from Eqs.~(\ref{eq-matrixG})-(\ref{eq-matrixF}), we can obtain the expression of $\mathbf{G}_{L}$. Here we only explicitly write out its diagonal elements:
\begin{eqnarray}&&G_{L11}=\frac{\bar{k}^2(\gamma_L+\gamma_R)+\gamma_R \Omega}{\bar{k}^2(\gamma_L+\gamma_R)^2+\gamma_L\gamma_R\Omega},\\
&&G_{L22}=\frac{\bar{k}^2(\gamma_L+\gamma_R)}{\bar{k}^2(\gamma_L+\gamma_R)^2+\gamma_L\gamma_R\Omega},
\end{eqnarray}
where $\Omega \equiv\left(k_1-k_2\right)^2+\left(\gamma_L+\gamma_R\right)\left(k_1 \gamma_R+k_2 \gamma_L\right)$.

Similarly, taking $\mathbf{E}_R=\left[\begin{array}{cc}0 & 0 \\ 0 & 1\end{array}\right]$, from Eqs.~(\ref{eq-matrixG})-(\ref{eq-matrixF}), we achieve the diagonal elements of $\mathbf{B}_{R}$:
\begin{eqnarray}
&&G_{R11}=\frac{\bar{k}^2(\gamma_L+\gamma_R)}{\bar{k}^2(\gamma_L+\gamma_R)^2+\gamma_L\gamma_R\Omega},\\
&&G_{R22}=\frac{\bar{k}^2(\gamma_L+\gamma_R)+\gamma_L \Omega}{\bar{k}^2(\gamma_L+\gamma_R)^2+\gamma_L\gamma_R\Omega}.
\end{eqnarray}

With these diagonal elements, we can calculate $C_L$ and $C_R$ by using Eqs.~(\ref{eq-weightfactor2}) and (\ref{eq-weightfactor3}). The corresponding expressions are as follows:
\begin{eqnarray}\label{eq-CLtwoscillator1}
&&C_L=\frac{2 \bar{k}^2 \gamma_L\left(\gamma_L+\gamma_R\right)+\gamma_L \gamma_R \Omega}{2 \bar{k}^2\left(\gamma_L+\gamma_R\right)^2+2 \gamma_L \gamma_R \Omega},\\
&&C_R=\frac{2 \bar{k}^2 \gamma_R\left(\gamma_L+\gamma_R\right)+\gamma_L \gamma_R \Omega}{2 \bar{k}^2\left(\gamma_L+\gamma_R\right)^2+2 \gamma_L \gamma_R\Omega}.\label{eq-CLtwoscillator2}
\end{eqnarray}
The corresponding effective temperature is
\begin{eqnarray}
T_e&=&\frac{2 \bar{k}^2 \gamma_L\left(\gamma_L+\gamma_R\right)+\gamma_L \gamma_R \Omega}{2 \bar{k}^2\left(\gamma_L+\gamma_R\right)^2+2 \gamma_L \gamma_R \Omega}T_L\\
&+&\frac{2 \bar{k}^2 \gamma_R\left(\gamma_L+\gamma_R\right)+\gamma_L \gamma_R \Omega}{2 \bar{k}^2\left(\gamma_L+\gamma_R\right)^2+2 \gamma_L \gamma_R\Omega}T_R.
\end{eqnarray}

In general, the weight factors $C_L$ and $C_R$ are unequal to each other. If $\gamma_L=\gamma_R$, the system is symmetrical coupling to both baths. In this situation, Eqs.~(\ref{eq-CLtwoscillator1}) and (\ref{eq-CLtwoscillator2}) lead to $C_L=C_R=1/2$ and then the effective temperature $T_e=\bar{T}\equiv (T_L+T_R)/2$, which is in good agreement with our intuition that the effective temperature of a harmonic chain equals to the mean temperature of two baths when the system simultaneously satisfies two symmetry conditions: (i) the interactions between oscillators in the chain are of left-right symmetry; (ii) two ends of the chain are symmetrically coupled to two baths. It is noted that the first condition holds automatically for a chain of two oscillators. Thus if and only if the second condition is satisfied, the effective temperature for a chain of two oscillators equals to the mean temperature of two baths.

\subsection{A chain of three harmonic oscillators with two ends coupled to two baths}
\begin{figure}[htp!]
\includegraphics[width=8cm]{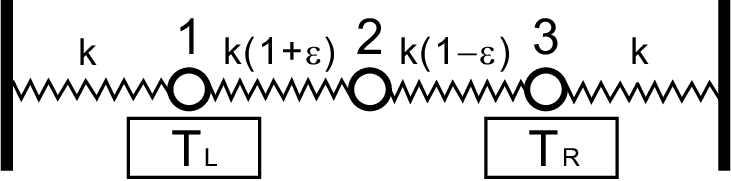}
\caption{A chain of three harmonic oscillators with two ends coupled to two baths at different temperatures $T_L$ and $T_R$. The elastic constants of four springs are $k$, $k(1+\varepsilon)$, $k(1-\varepsilon)$, and $k$, respectively.}\label{fig-threeoscillator}
\end{figure}

Let us consider a chain of three harmonic oscillators as shown in Fig.~\ref{fig-threeoscillator}. Two ends of the chain are coupled to two baths at different temperatures $T_L$ and $T_R$. The elastic constants of four springs are $k$, $k(1+\varepsilon)$, $k(1-\varepsilon)$, and $k$, respectively. Here $\varepsilon$ takes value in the domain between $-1$ and $1$, which represents the degree of left-right symmetry breaking in interactions between harmonic oscillators. In particular, $\varepsilon=0$ corresponds to a uniform chain with left-right symmetry where four springs possess the same elastic constants.

The elastic matrix can be expressed as
\begin{equation}\label{eq-elasticmatrix3}\mathbf{K}=k\left[\begin{array}{ccc}
2+\varepsilon & -(1+\varepsilon) & 0 \\
-(1+\varepsilon) & 2 & -(1-\varepsilon) \\
0 & -(1-\varepsilon) & 2-\varepsilon
\end{array}\right].
\end{equation}
For simplicity, the reference elastic constant $k$ is set to be unity in the following discussions. The frictional matrix is taken as \begin{equation}\label{eq-frictmatrix3}\bm\Gamma=\gamma\left[\begin{array}{lll}1+\delta & 0 & 0 \\ 0 & 0 & 0 \\ 0 & 0 & 1-\delta\end{array}\right],\end{equation}
where $\gamma$ is the reference frictional constant which is set to be unity in the following discussions. The parameter $\delta$ takes value in the domain between $-1$ and $1$, which represents the degree of asymmetry of coupling between the chain and two baths. In particular, $\delta=0$ corresponds to the situation of symmetric coupling between the chain and two baths.
We will discuss the influences of $\varepsilon$ and $\delta$ on the effective temperature in details.

Taking $\mathbf{E}_L=\left[\begin{array}{ccc}1 & 0 & 0\\ 0&0&0\\ 0&0 & 0\end{array}\right]$ and $\mathbf{E}_R=\left[\begin{array}{ccc}0 & 0 & 0\\ 0&0&0\\ 0&0 & 1\end{array}\right]$, we can obtain the expressions of $\mathbf{G}_{L}$ and $\mathbf{G}_{R}$ respectively from Eqs.~(\ref{eq-matrixG})-(\ref{eq-matrixF}). Substituting the traces of $\mathbf{G}_{L}$ and $\mathbf{G}_{R}$ into Eqs.~(\ref{eq-weightfactor2}), we finally obtain the weight factor
\begin{eqnarray}
C_L &=&\frac{1}{3}+\frac{u_{22}(1-\varepsilon^2)^2(1- 4 \delta-\delta^2)}{6 \Delta(1-\delta^2)}\nonumber \\
&+&\frac{(u_{23}^2-u_{22}u_{33})(1+\varepsilon)^2+u_{12}u_{23}(1-\varepsilon^2)}{3 \Delta},\label{eq-CLespilondelta}
\end{eqnarray}
with $u_{11}=(1-\varepsilon)^2/2+[{\varepsilon^2+(1+\varepsilon)^2}]/({1+\delta})+2(1+\delta)$, $u_{12}=(1-\varepsilon)[\varepsilon+{\varepsilon}/({1+\delta})-(1+\delta)]$, $u_{22}=4+2 \varepsilon^2+(1+\varepsilon)^2/(1-\delta)-2 \varepsilon \delta+(1-\varepsilon)^2/(1+\delta)$, $u_{23}=-(1+\varepsilon)[\varepsilon+{\varepsilon}/({1-\delta})+(1-\delta)]$, $u_{33}=(1+\varepsilon)^2/2+[(1-\varepsilon)^2+\varepsilon^2]/({1-\delta})+2(1-\delta)$, and $\Delta=[(u_{12}(1-\varepsilon)+u_{23}(1+\varepsilon)]^2-u_{22}[a_{11}(1-\varepsilon)^2+u_{33}(1+\varepsilon)^2-(1-\varepsilon^2)^2]$. 

\begin{figure}[htp!]
\includegraphics[width=8cm]{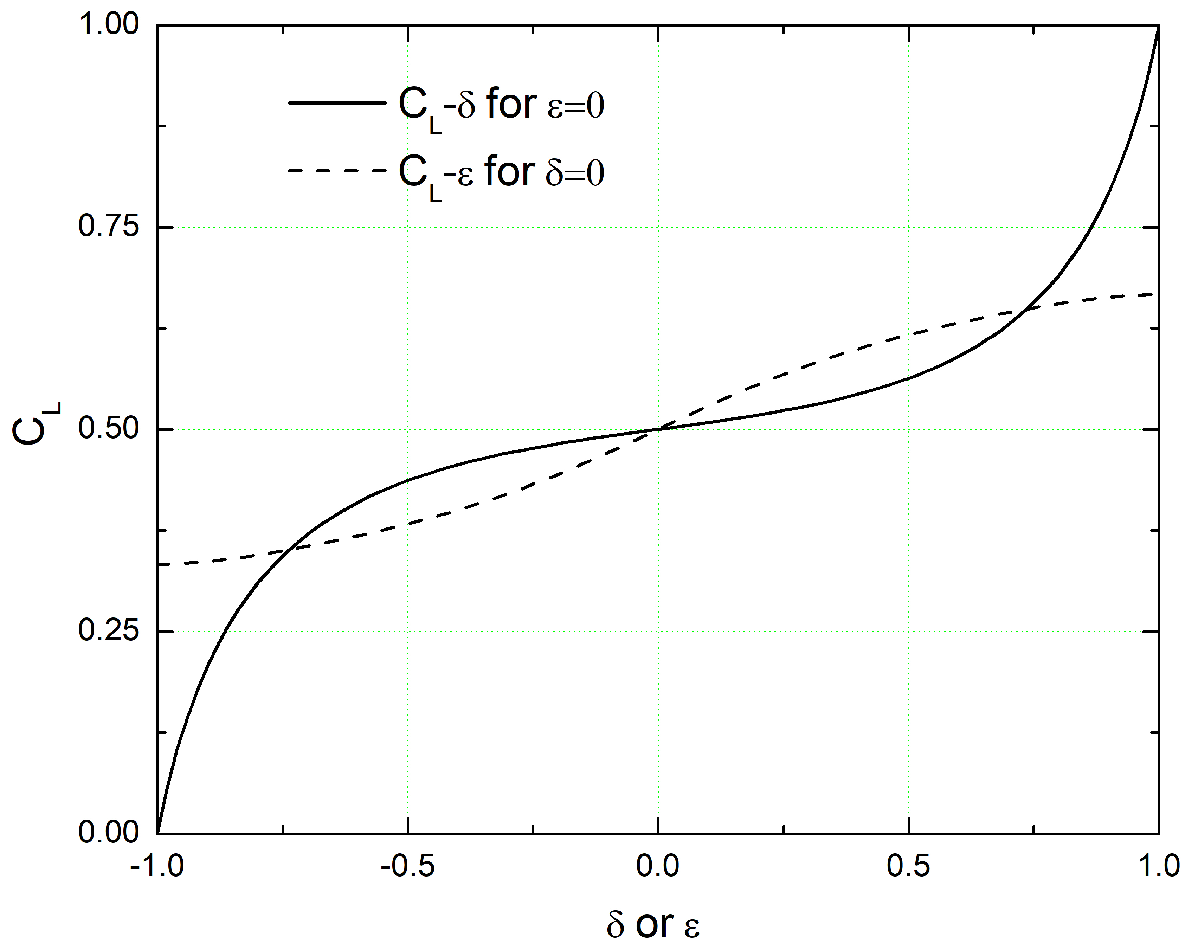}
\caption{(Color online) Graph of functions (\ref{eq-CL-delta}) and (\ref{eq-CL-epsilon}).}\label{fig-CL-deltaoreps}
\end{figure}

The expression of weight factor, Eq.~(\ref{eq-CLespilondelta}), is too complicated to be see clearly the dependence of $C_L$ on $\varepsilon$ and $\delta$. We may gain an intuitive understanding on the behavior of $C_L$ by presetting $\varepsilon=0$ or $\delta=0$. In the situation of $\varepsilon=0$, the above equation~(\ref{eq-CLespilondelta}) is reduced to
\begin{equation}\label{eq-CL-delta}
C_L=\frac{24+4\delta-30 \delta^2+2 \delta^3+9 \delta^4-3 \delta^5}{48-60 \delta^2+18 \delta^4}.
\end{equation}
The graph of function (\ref{eq-CL-delta}) is depicted as solid line in Fig.~\ref{fig-CL-deltaoreps}. 
We observe that $C_L=0$ for $\delta=-1$ and $C_L=1$ for $\delta=1$, and that $C_L$ monotonically increases with $\delta$. These behaviors are consistent with our intuition.
On the one hand, we can obtain $T_e=T_R$ from Eq.~(\ref{eq-effectivetemp}) with considering $C_L=0$ and $C_R=1-C_L=1$ for $\delta=-1$.
On the other hand, frictional matrix (\ref{eq-frictmatrix3}) with $\delta=-1$ implies that the chain is strongly coupled with the bath at temperature $T_R$ but decoupled with the bath at temperature $T_L$. 
Undoubtedly, the effective temperature should be equal to $T_R$. 
When the value of $\delta$ increases, the coupling between the chain and the bath at temperature $T_L$ is enhanced, while the coupling between the chain and the bath at temperature $T_R$ is weakened. Thus the weight of contribution of $T_L$ in the effective temperature rises, which exactly corresponds to the monotonic increase of $C_L$ with $\delta$. When $\delta=1$, the chain is strongly coupled with the bath at temperature $T_L$ but decoupled with the bath at temperature $T_R$. Undoubtedly, the effective temperature should be equal to $T_L$ which corresponds to $C_L=1$.

Similarly, Eq.~(\ref{eq-CLespilondelta}) is degenerated into
\begin{equation}\label{eq-CL-epsilon}
C_L=\frac{24+14 \varepsilon+33 \varepsilon^2+2 \varepsilon^3-27 \varepsilon^4-2 \varepsilon^5+12 \varepsilon^6}{48+66 \varepsilon^2-54 \varepsilon^4+24 \varepsilon^6}
\end{equation}
when $\delta=0$.
The graph of function (\ref{eq-CL-epsilon}) is depicted as dashed line in Fig.~\ref{fig-CL-deltaoreps}. 
We find that $C_L=1/3$ for $\varepsilon=-1$ and $C_L=2/3$ for $\varepsilon=1$, and that $C_L$ monotonically increases with $\varepsilon$. These behaviors are also consistent with our intuition.
On the one hand, we can obtain $T_e=T_L/3+2T_R/3$ from Eq.~(\ref{eq-effectivetemp}) with considering $C_L=1/3$ and $C_R=1-C_L=2/3$ for $\varepsilon=-1$.
On the other hand, elastic matrix (\ref{eq-elasticmatrix3}) with $\varepsilon=-1$ implies that the interaction between oscillators 1 and 2 is vanishing. Thus oscillator 1 and the system consisting of oscillators 2 and 3 are two independent subsystems. These two subsystems are in equilibrium with the baths at temperatures $T_L$ and $T_R$, respectively. Thus the weight of contribution of $T_L$ in the effective temperature is $1/(1+2)=1/3$. Similar interpretation holds for the situation of $\varepsilon=1$. 
In addition, with the increase of $\varepsilon$, the interaction between oscillators 1 and 2 is enhanced, while interaction between oscillators 2 and 3 is weakened. Thus the weight of influence of $T_L$ to the effective temperature rises, which corresponds to the monotonic increase of $C_L$ with $\varepsilon$.

\begin{figure}[htp!]
\includegraphics[width=8cm]{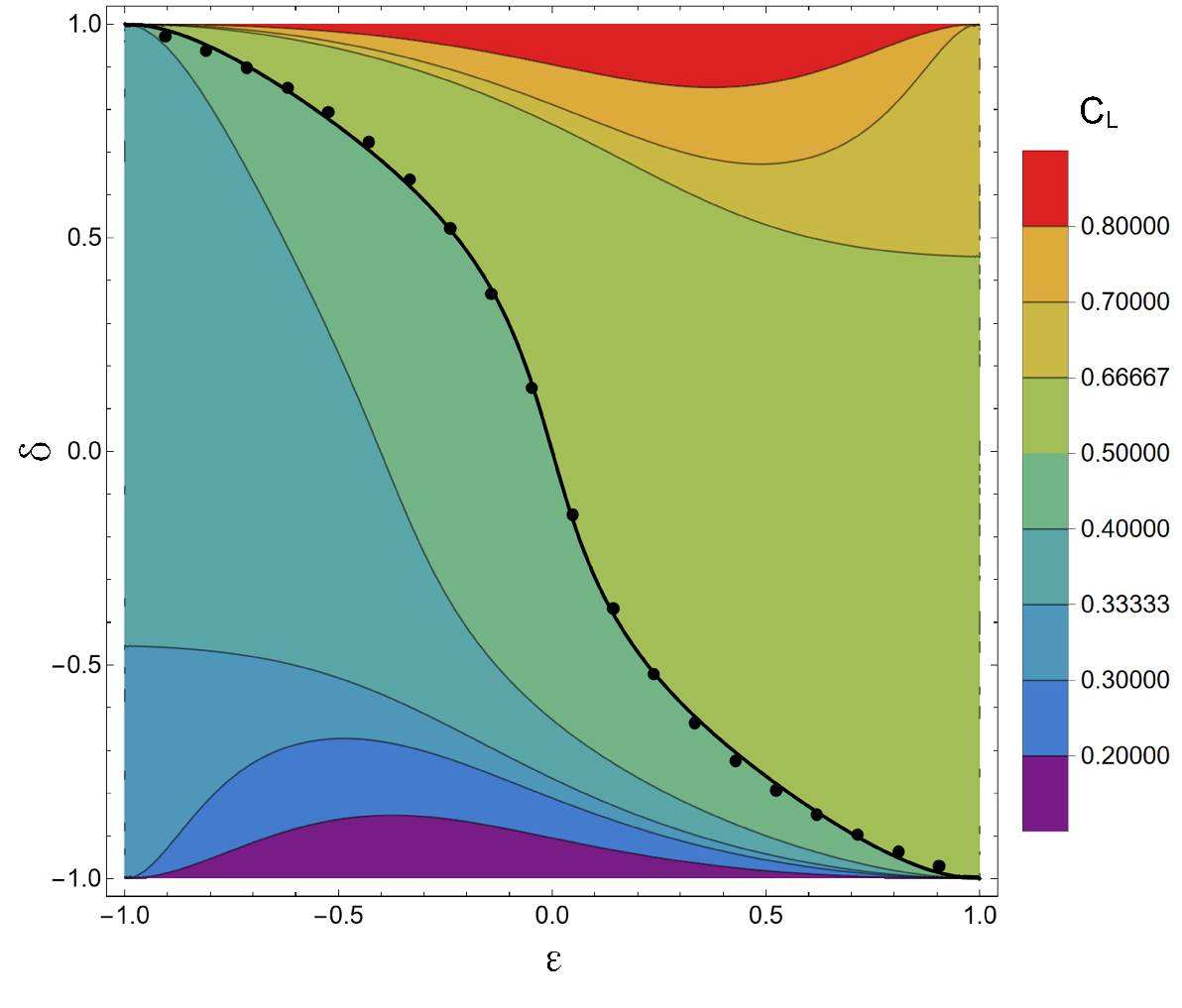}
\caption{(Color online) Contour plot of function (\ref{eq-CLespilondelta}). The thick solid line is the contour line corresponding to $C_L=1/2$. This line can be approximately represented by a function $\delta\approx -{7 \varepsilon}/(2+5 |\varepsilon|)$ (Dots in the graph).}\label{fig-CLespilondelta}
\end{figure}

Furthermore, we draw a contour plot of function (\ref{eq-CLespilondelta}) in Fig.~\ref{fig-CLespilondelta} to understand the complicated dependency behavior of $C_L$ on $\varepsilon$ and $\delta$. At a glance, we see that the graph is invariant under rotating $180^\circ$ around the centre $(0,0)$ provided that replacing $C_L$ with $C_R=1-C_L$. The underlying reason is that the system is in fact unchanged if we merely reassign labels 1, 2, and 3 of the oscillators as well as labels $L$ and $R$ of two baths. Detailed survey manifests a subtle difference between the behavior of $\delta-$dependence of $C_L$ for given $\varepsilon$ and that of $\varepsilon-$dependence for given $\delta$. For given $\varepsilon$, $C_L$ is a monotonically increasing function of $\delta$. However, $C_L$ is not always a monotonically increasing function of $\varepsilon$ for given $\delta$. Only if $|\delta|$ is not too large, for example smaller than 0.4, $C_L$ is a monotonically increasing function of $\varepsilon$. Otherwise, $C_L$ is not a monotonic function of $\varepsilon$. It is easy to understand that $C_L$ is a monotonically increasing function of $\delta$. However, it is not straightforward for us to interpret the surprising nonmonotonic dependence of $C_L$ on $\varepsilon$. The root of this behavior might be a transition experienced by some oscillators from underdamping to overdamping. 

Particularly, from Fig.~\ref{fig-CLespilondelta} we observe that $C_L=1/2$ when $\varepsilon$ and $\delta$ are simultaneously vanishing, which is in good agreement with our intuition that the effective temperature of the harmonic chain equals to the mean temperature of two
baths when the system simultaneously satisfies two symmetry conditions mentioned at the end of Sec.~\ref{subsec-twooscillators}. We should point out that this two symmetry conditions are just sufficient conditions but not necessary conditions since the increasing of $\varepsilon$ and decreasing of $\delta$ contribute the opposite effects in $C_L$. We specifically draw the contour line corresponding to $C_L=1/2$, the thick solid line in Fig.~\ref{fig-CLespilondelta}. The explicit relation between $\varepsilon$ and $\delta$ is too complicated, which may be approximately expressed as $\delta\approx -{7 \varepsilon}/(2+5 |\varepsilon|)$.

\section{Conclusion and discussion\label{sec-summary}}
We have investigated the effective temperature at steady state of a harmonic chain coupled to two thermal baths at different temperatures.
The key ansatz is that the revised covariance matrix may be decomposed into the diagonal matrix and traceless residual matrix [see Eqs.~(\ref{eq-effecttemp1}) and (\ref{eq-traceless})]. 
We suggest taking the weighted average temperature [Eq.~(\ref{eq-effectivetemp})] as the effective temperature of the system. The weight factors [Eq.~(\ref{eq-weightfactor})] are related to the coupling constants between the system and two baths as well as the asymmetry of interactions between oscillators. The residual matrix [Eq.~(\ref{eq-sigma-r})] depends linearly on the temperature difference between two thermal baths. 
The advantage of weighted average temperature lies in revisiting the thermodynamics of nonequilbrium steady states. The fundamental thermodynamic relations in nonequilbrium steady states such as internal energy [Eq.~(\ref{eq-intenergy2})], entropy [Eq.~(\ref{eq-entropy})], and free energy [Eq.~(\ref{eq-freeenergy})] possess similar concise forms as those in equilibrium thermodynamics. The minor difference is to replace the temperature in equilibrium with the weighted average temperature in steady states. The nonequilbrium character manifests in the heat transfer and entropy production. The heat transfer rate [Eq.~(\ref{eq-heattrans2})] is proportional to the temperature difference between two baths, while the entropy production rate [Eq.~(\ref{eq-entropyprodrate2})] is proportional to the quadratic term of temperature difference. Finally, we also illustrate the procedure to calculate the effective temperatures via three examples. 

Before ending this paper, we address several open issues which will be investigated in the future work.

1). The model that we employed is a linear system. If the equations of motion contain nonlinear forces, the steady-state distribution cannot directly be expressed with the covariance matrix. Therefore, our definition of effective temperature based on the decomposition of covariance matrix is inapplicable to a nonlinear system directly. A straightforward extension might be considering the linearization of the nonlinear system and then defining the effective temperature of the corresponding linearized system. The other challenging way might be extending the concept of covariance matrix to the nonlinear system. A candidate matrix might be $\langle\mathbf{z}(\partial H/\partial \mathbf{z})\rangle_{t\rightarrow \infty}$ where $\mathbf{z}=[x_1,\cdots,x_N,p_1,\cdots,p_N]^\mathrm{T}$ and $H$ is the Hamiltonian of the nonlinear system. $\partial H/\partial \mathbf{z}$ is the abbreviation of $[\partial H/\partial x_1,\cdots,\partial H/\partial x_N,\partial H/\partial p_1,\cdots,\partial H/\partial p_N]$.

2). We have proposed to define the effective temperature via the weighted average temperature. For our model system, this definition is consistent to the kinetic temperature in literature (see also the discussion at the end of Sec.~\ref{subsec-intenergy}). In fact, another typical proposal to define effective temperature for a system driven away from equilibrium under an external field (usually not a temperature difference) is based on the generalized linear fluctuation-dissipation relation~\cite{Jou2003,Cugliandolo1997,Cugliandolo2011,Puglisi2017,DiVentra2019,Fielding2002,Jabbari-Farouji2008,Ben-Isaac2011}. It is unclear to make a connection between two kinds of definitions for a system coupled to two baths at different temperature. We expect that fluctuation-dissipation relation keeps its conventional form for small enough temperature difference $\Delta T\equiv T_L-T_R$. However, this is a trivial case $T_e\approx T_L\approx  T_R$. We are more interested in the situation of large temperature difference where the fluctuation-dissipation relation should be replaced by the fluctuation theorem for heat exchange~\cite{Jarzynski230602,Saito180601,Lahiri2014,Quan024135}.

3). As case study, we have analytically investigated harmonic chains of $N=1,~2,~3$ oscillators. In general, the chains is inhomogeneous when $N\ge 2$. The results support our conjecture that the effective temperature of a harmonic chain equals to the mean temperature of two baths when the system simultaneously satisfies two symmetry conditions: (i) the interactions between oscillators in the chain are of left-right symmetry; (ii) two ends of the chain are symmetrically coupled to two baths. Rieder~\textit{et al.} found that the kinetic temperature of a long homogeneous chain coupled symmetrically to two baths is almost the mean temperature of two baths~\cite{Rieder-Lebowitz1967}, which is consistent with our conjecture. Kannan~\textit{et al.} investigated a harmonic crystal with alternating masses, and their numerical results reveal that the local kinetic temperature profile oscillates with period two in the bulk of the system~\cite{KannanDharPRE2012}. In a special case of odd $N$, the local temperature in the bulk is found to have uniform profile with value $(T_L+T_R)/2$~\cite{KannanDharPRE2012}. This result also supports our conjecture since two symmetry conditions mentioned above hold exactly for this special case of odd $N$ in their model. There seems no apparent obstacle to establishing an explicit connection between our definition of effective temperature and the profile of local temperature obtained by Kannan~\textit{et al.} in the absence of symmetry. In a more complicated case of disordered harmonic chain with infinite length~\cite{Matsuda-IshiiPTPS1970}, it might be a challenge to obtain an explicit expression of effective temperature.

4). Although we have merely confined our research object to a thermal system in contact with to two baths. The idea of weighted average temperature might provide some insight to athermal granular systems. Due to breakdown of energy equipartition, it is found that particles of different mass have different granular temperatures in a muticomponent granular gas~\cite{Losert-Chaos682,Menon-PRL198301,IppolitoPRE2017,WangPRE031301}. Averaging these different granular temperatures with weight factors being the masses of different types of particles, one might be able to define an effective temperature to characterize the behavior of multicomponent granular gas. It should be noted that another definition of ``temperature" (compactivity) for jammed granular packings was proposed by Edwards and co-workers~\cite{Edwards1080,Edwards1091} who introduced a volume ensemble of equiprobable jammed states in analogy to equilibrium statistical physics. The Edwards temperature has attracted much attention by both theoretical and experimental physicists~\cite{Baule015006}. Particularly, recent experimental results not only support the validity of the Edwards volume ensemble, but also demonstrate the equivalence between the Edwards temperature and the temperature defined according to the fluctuation-dissipation relation~\cite{WangYJ018002,WangYJ228004}. It is a difficult task to establish a connection between our definition of effective temperature and the Edwards temperature.

5). Our discussions are focused on classical thermodynamic systems. The quantum dissipative systems and steady-state transports have been discussed deeply \cite{Ness2014,CaoJ2019,CaldeiraAP1983,WeissBook2021,CaldeiraBook2014,LandiRMP045006,Yan-Shao2016,TalknerPRA3278,Carlen240806887,WeiderpassPRB125401,DharPRB195405,DharPRE011126,NetoPRE031116,AsadianPRE012109,OliveiraPRE032139,HuBLAN2015}.
We note that the covariance matrix has also been used to investigate the quantum transport in harmonic chains~\cite{TalknerPRA3278,DharPRE011126,OliveiraPRE032139,LandiRMP045006}. It should be straightforward for us to extend our discussions based on the covariance matrix from classical situation to quantum realm. We believe that the idea of weighted average temperature holds also for a quantum system coupled to two baths.

\section*{Acknowledgement}
The author thanks Jin Wang (Stony Brook University) who attracts the author's attention to the present topic in this work. The author thanks Ning Xu (University of Science and Technology of China) and Yujie Wang (Shanghai Jiao Tong University, and Chengdu University of Technology) for their kind helps.
The author is also grateful for the financial support from the National Natural Science Foundation of China (Grant No. 11975050).

\appendix

\section{Derivation of Eq.~(\ref{eq-resmatrix})\label{sec-ap-12to13}}
Here we will derive Eq.~(\ref{eq-resmatrix}) in details. 

Since both $\tilde{\mathbf{K}}$ and $\bm{\sigma}$ are symmetric matrices, then we have $\tilde{\bm{\sigma}}^\mathrm{T}=(\bm{\sigma}\tilde{\mathbf{K}})^\mathrm{T}=\tilde{\mathbf{K}}\bm{\sigma}$ which is in general unequal to $\tilde{\bm{\sigma}}$. Similarly, we have $\tilde{\mathbf{A}}^T=(\tilde{\mathbf{K}}\mathbf{A})^\mathrm{T}={\mathbf{A}}^\mathrm{T}\tilde{\mathbf{K}}$. Note that $\mathbf{A}$ is usually not a symmetric matrix. Multiplying $\tilde{\mathbf{K}}$ left and right towards Eq.~(\ref{eq-Lyapunov}), we have
\begin{equation}\tilde{\mathbf{A}}\tilde{\bm{\sigma}}+\tilde{\bm{\sigma}}^\mathrm{T}\tilde{\mathbf{A}}^\mathrm{T}=2\mathbf{D}_\mathrm{t},\label{eq-Lyapunovtilde}
\end{equation}
where we have used the fact $\tilde{\mathbf{K}}\mathbf{D}_\mathrm{t}\tilde{\mathbf{K}}=\mathbf{D}_\mathrm{t}$ which can be directly confirmed by the definition of $\tilde{\mathbf{K}}$ and $\mathbf{D}_\mathrm{t}$.

In addition, we obtain $\tilde{\mathbf{A}}+\tilde{\mathbf{A}}^T=2\left[\begin{array}{cc}\mathbf{0} & \mathbf{0} \\ \mathbf{0} &\bm{\Gamma}\end{array}\right]$
from Eq.~(\ref{eq-tildeAmatrx}). This relation implies
\begin{equation}\label{eq-AplusAt}
\tilde{\mathbf{A}}(T_e\mathbf{I}_{2N})+(T_e\mathbf{I}_{2N})\tilde{\mathbf{A}}^T=2\left[\begin{array}{cc}\mathbf{0} & \mathbf{0} \\ \mathbf{0} & T_e\bm{\Gamma}\end{array}\right]
\end{equation}
since $\mathbf{I}_{2N}$ is the identity matrix.
Subtracting Eqs.~(\ref{eq-Lyapunovtilde}) and (\ref{eq-AplusAt}), and considering Eq.~(\ref{eq-effecttemp1}), we will soon arrive at Eq.~(\ref{eq-resmatrix}).

\section{Traces of $\mathbf{B}_\alpha$ and $\mathbf{G}_\alpha$ \label{equaltraceBG}}
Here we will discuss the relationship between the traces of $\mathbf{B}_\alpha$ and $\mathbf{G}_\alpha$.

First, we will prove $\mathbf{J}_\alpha^\mathrm{T}=-\mathbf{J}_\alpha$.
From Eq.~(\ref{eq-matrixFJ}), we have $\mathbf{J}_\alpha^\mathrm{T}=\mathbf{F}_\alpha\mathbf{K}^{-1}$ and then $\mathbf{J}_\alpha=\mathbf{K}^{-1}\mathbf{F}_\alpha^\mathrm{T}$ with the consideration of $\mathbf{K}^\mathrm{T}=\mathbf{K}$. From Eq.~(\ref{eq-matrixKF}), we have $\mathbf{K}^{-1}\mathbf{F}_\alpha^\mathrm{T}=-\mathbf{F}_\alpha\mathbf{K}^{-1}$. It follows that $\mathbf{J}$ is an antisymmetric matrix:
\begin{equation}\label{eq-Jantisymmetric}
\mathbf{J}_\alpha^\mathrm{T}=-\mathbf{J}_\alpha.
\end{equation}

Second, we will prove $\mathrm{Tr}(\bm{\Gamma}_\alpha\mathbf{J}_\alpha^\mathrm{T})=0$.
The antisymmetry (\ref{eq-Jantisymmetric}) implies $\mathrm{Tr}(\bm{\Gamma}_\alpha\mathbf{J}_\alpha^\mathrm{T})=\mathrm{Tr}(-\bm{\Gamma}_\alpha\mathbf{J}_\alpha)=-\mathrm{Tr}(\bm{\Gamma}_\alpha\mathbf{J}_\alpha)$.
Since a matrix and its transpose have the same traces, we obtain $\mathrm{Tr}(\bm{\Gamma}_\alpha\mathbf{J}_\alpha^\mathrm{T})=\mathrm{Tr}(\bm{\Gamma}_\alpha\mathbf{J}_\alpha^\mathrm{T})^\mathrm{T}=\mathrm{Tr}(\mathbf{J}_\alpha\bm{\Gamma}_\alpha^\mathrm{T})$.
Considering the symmetry of $\bm{\Gamma}_\alpha$ and the commutativity of trace of matrix product, we further arrive at $\mathrm{Tr}(\bm{\Gamma}_\alpha\mathbf{J}_\alpha^\mathrm{T})=\mathrm{Tr}(\bm{\Gamma}_\alpha\mathbf{J}_\alpha)$. 
Therefore, the only possible consequence is
\begin{equation}\label{eq-traceGammaJ}
\mathrm{Tr}(\bm{\Gamma}_\alpha\mathbf{J}_\alpha^\mathrm{T})=0.
\end{equation}

Finally, we will prove $\mathrm{Tr}\mathbf{B}_\alpha=\mathrm{Tr}\mathbf{G}_\alpha$. From Eq.~(\ref{eq-matrixKBG}) we have $\mathbf{B}_\alpha=\mathbf{K}^{-1}\mathbf{G}_\alpha \mathbf{K}-\mathbf{K}^{-1}\bm{\Gamma}\mathbf{F}_\alpha$ which implies $\mathrm{Tr}\mathbf{B}_\alpha=\mathrm{Tr}(\mathbf{K}^{-1}\mathbf{G}_\alpha \mathbf{K})-\mathrm{Tr}(\mathbf{K}^{-1}\bm{\Gamma}\mathbf{F}_\alpha)$. Since the trace of matrix product is invariant under commutating two matrices, we deduce $\mathrm{Tr}(\mathbf{K}^{-1}\mathbf{G}_\alpha \mathbf{K})=\mathrm{Tr}(\mathbf{G}_\alpha \mathbf{K}\mathbf{K}^{-1})=\mathrm{Tr}\mathbf{G}_\alpha$ and $\mathrm{Tr}(\mathbf{K}^{-1}\bm{\Gamma}\mathbf{F}_\alpha)=\mathrm{Tr}(\bm{\Gamma}\mathbf{F}_\alpha\mathbf{K}^{-1})$. Considering $\mathbf{F}_\alpha\mathbf{K}^{-1}=\mathbf{J}_\alpha^\mathrm{T}$ and Eq.~(\ref{eq-traceGammaJ}), we further obtain $\mathrm{Tr}(\mathbf{K}^{-1}\bm{\Gamma}\mathbf{F}_\alpha)=\mathrm{Tr}(\bm{\Gamma}_\alpha\mathbf{J}_\alpha^\mathrm{T})=0$. Therefore, we find that the traces of $\mathbf{B}_\alpha$ and $\mathbf{G}_\alpha$ are identical:
\begin{equation}\label{eq-traceBandG}
\mathrm{Tr}\mathbf{B}_\alpha=\mathrm{Tr}\mathbf{G}_\alpha.
\end{equation}

\section{Procedure to calculate $C_L$ for the chain of three oscillators\label{procedure3CL}}
It is technically involved to directly solve matrix equations~(\ref{eq-matrixG})-(\ref{eq-matrixF}) for an inhomogeneous chain consisting of $N>2$ oscillators. Since our aim is to calculate $C_L$, here we may provide a relatively simplified procedure to obtain the elements of matrix $\mathbf{B}_\alpha$ for $N=3$. Using this procedure, we may write the final expression of $C_L$ in a compact form (\ref{eq-CLespilondelta}) that makes us not too dizzy.

From Eqs.~(\ref{eq-matrixG})-(\ref{eq-matrixKBG}), we derive
\begin{equation}\label{eq-KG-GK}
\mathbf{K} \mathbf{G}_\alpha-\mathbf{G}_\alpha \mathbf{K}=\bm{\Gamma} \mathbf{J}_\alpha\mathbf{K}+\mathbf{K J}_\alpha \bm{\Gamma},
\end{equation}
and
\begin{equation}\label{eq-KJ-JK-G}
\mathbf{K}\mathbf{J}_\alpha-\mathbf{J}_\alpha\mathbf{K}+\bm{\Gamma}\mathbf{G}_\alpha+\mathbf{G}_\alpha\bm{\Gamma}=2\mathbf{E}_\alpha.
\end{equation}
The independent elements of antisymmetric matrix $\mathbf{J}_\alpha$ and symmetric matrix $\mathbf{G}_\alpha$ are denoted as $J_{12}$, $J_{13}$, $J_{23}$, $G_{11}$, $G_{12}$, $G_{13}$, $G_{22}$, $G_{23}$, $G_{33}$, respectively. Introducing two vectors $\mathbf{v}=[J_{12}, J_{13}, J_{23},G_{22}]^\mathrm{T}$ and $\mathbf{g}=[G_{11},G_{12}, G_{13}, G_{23}, G_{33}]^\mathrm{T}$, from Eq.~(\ref{eq-KG-GK}) and (\ref{eq-KJ-JK-G}) we derive
\begin{eqnarray}
&&\mathbf{U}_2 \mathbf{g} = \mathbf{U}_3 \mathbf{v}\label{eq-u2gu3v},\\
&&\mathbf{g} = \mathbf{w}_\alpha- \mathbf{U}_1 \mathbf{v}\label{eq-gwu1v}.
\end{eqnarray}
We have considered elastic matrix (\ref{eq-elasticmatrix3}) and frictional matrix (\ref{eq-frictmatrix3}) in derivation of these two equations.
Here three matrices $\mathbf{U}_1$, $\mathbf{U}_2$ and $\mathbf{U}_3$ are explicitly expressed as
\begin{equation}\label{eq-matrixU1}
\mathbf{U}_1=\left[\begin{array}{cccc}
\frac{1+\varepsilon}{1+\delta} & 0 & 0 & 0 \\
\frac{\varepsilon}{1+\delta} & \frac{1-\varepsilon}{1+\delta} & 0 & 0 \\
\frac{1-\varepsilon}{2} & \varepsilon & -\frac{1+\varepsilon}{2} & 0 \\
0 & -\frac{1+\varepsilon}{1-\delta} & \frac{\varepsilon}{1-\delta} & 0 \\
0 & 0 & -\frac{1-\varepsilon}{1-\delta} & 0
\end{array}\right],
\end{equation}
\begin{equation}\label{eq-matrixU2}
\mathbf{U}_2=\left[\begin{array}{ccccc}
1+\varepsilon & \varepsilon & 1-\varepsilon & 0 & 0 \\
0 & 1-\varepsilon & 2 \varepsilon & -1-\varepsilon & 0 \\
0 & 0 & -1-\varepsilon & \varepsilon & -1+\varepsilon \\
0 & 0 & 0 & 0 & 0
\end{array}\right],
\end{equation}
and\begin{widetext}
\begin{equation}\label{eq-matrixU3}
\mathbf{U}_3=\left[\begin{array}{cccc}
2(1+\delta) & -(1-\varepsilon)(1+\delta) & 0 & 1+\varepsilon \\
-(1-\varepsilon)(1+\delta) & 4-2 \varepsilon \delta & -(1+\varepsilon)(1-\delta) & 0 \\
0 & -(1+\varepsilon)(1-\delta) & 2(1-\delta) & -1+\varepsilon \\
1+\varepsilon & 0 & -1+\varepsilon & 0
\end{array}\right]
.\end{equation}\end{widetext}
With these matrices, we can solve $\mathbf{v}$ and $\mathbf{g}$ from Eqs.~(\ref{eq-u2gu3v}) and (\ref{eq-gwu1v}). Their formal expressions are as follows:
\begin{eqnarray}
\mathbf{v}&=&\mathbf{U}^{-1}\mathbf{U}_2\mathbf{w}_\alpha,\label{eq-solut-v}\\
\mathbf{g}&=&\mathbf{w}_\alpha-\mathbf{U}_1\mathbf{U}^{-1}\mathbf{U}_2\mathbf{w}_\alpha. \label{eq-solut-g}
\end{eqnarray}
where $\mathbf{U}\equiv\mathbf{U}_2\mathbf{U}_1+\mathbf{U}_3$ is expressed in matrix form:
\begin{equation}\label{eq-matrix-Utot}
\mathbf{U}=\left[\begin{array}{cccc}
u_{11} & u_{12} & -\frac{1-\varepsilon^2}{2} & 1+\varepsilon \\
u_{12} & u_{22} & u_{23} & 0 \\
-\frac{1-\varepsilon^2}{2} & u_{23} & u_{33} & -(1-\varepsilon) \\
1+\varepsilon & 0 & -(1-\varepsilon) & 0
\end{array}\right].
\end{equation}
The elements $u_{11}$, $u_{12}$, $u_{22}$, $u_{23}$, $u_{33}$ and the determinant of $\mathbf{U}$ have been presented below Eqs.~(\ref{eq-CLespilondelta}).

Vectors $\mathbf{w}_\alpha$ ($\alpha=L,~R$) in Eqs.~(\ref{eq-gwu1v}), (\ref{eq-solut-v}) and (\ref{eq-solut-g}) depend on $\mathbf{E}_\alpha$, which may be explicitly expressed as
$\mathbf{w}_L=[{1}/({1+\delta}), 0,0,0,0]^\mathrm{T}$ and $\mathbf{w}_R=[0,0,0,0,{1}/({1-\delta})]^\mathrm{T}$. Substituting $\mathbf{w}_L$ into Eqs.~(\ref{eq-solut-v}) and (\ref{eq-solut-g}), we can obtain the diagonal elements of matrix $\mathbf{G}_L$:
\begin{eqnarray}
G_{L11}&=&\frac{1}{1+\delta}+\frac{u_{22}(1-\varepsilon^2)^2}{\Delta(1+\delta)^2} \\
G_{L22}&=&\frac{u_{22}(1-\varepsilon^2)^2}{2\Delta(1+\delta)}+\frac{u_{12}u_{23}(1-\varepsilon^2)}{\Delta(1+\delta)}\nonumber \\
&&+\frac{(u_{23}^2-u_{22}u_{33})(1+\varepsilon)^2}{\Delta(1+\delta)}\\
G_{L33}&=&-\frac{u_{22}(1-\varepsilon^2)^2}{\Delta(1-\delta^2)}
\end{eqnarray}
Similarly, substituting $\mathbf{w}_R$ into Eqs.~(\ref{eq-solut-v}) and (\ref{eq-solut-g}), we can obtain the diagonal elements of matrix $\mathbf{G}_R$:
\begin{eqnarray}
G_{R11}&=&-\frac{u_{22}(1-\varepsilon^2)^2}{\Delta(1-\delta^2)}\\
G_{R22}&=&\frac{u_{22}(1-\varepsilon^2)^2}{2\Delta(1-\delta)}+\frac{u_{12}u_{23}(1-\varepsilon^2)}{\Delta(1-\delta)}\nonumber \\
&&+\frac{(u_{12}^2-u_{11}u_{22})(1-\varepsilon)^2}{\Delta(1-\delta)}\\
G_{R33}&=&\frac{1}{1-\delta}+\frac{u_{22}(1-\varepsilon^2)^2}{\Delta(1-\delta)^2} 
\end{eqnarray}
Considering $\gamma_L=1+\delta$, $\gamma_R=1-\delta$, $\mathrm{Tr}\mathbf{G}_L=G_{L11}+G_{L22}+G_{L33}$, $\mathrm{Tr}\mathbf{G}_R=G_{R11}+G_{R22}+G_{R33}$, we can derive Eq.~(\ref{eq-CLespilondelta}) from definition (\ref{eq-weightfactor2}) of $C_L$.

\section{Revisit the asymmetrical chain investigated by Wu and Wang}
Let us consider a chain of three harmonic oscillators with two ends coupled to two baths. We adopt the same values of elastic matrix and frictional matrix as those in Ref.~\cite{Wu-Wang2022}. For $N=3$, the elastic matrix is
$\mathbf{K}=k\left[\begin{array}{ccc}
2 & -1 & 0 \\
-1 & 3 & -2 \\
0 & -2 & 2
\end{array}\right],$
where $k$ is the reference elastic constant which is set to be unity in the following discussions. The off-diagonal elements of $\mathbf{K}$ are different from each other, which implies that interactions between harmonic oscillators are asymmetrical. The frictional matrix is $\bm\Gamma=\gamma\left[\begin{array}{ccc}1 & 0 & 0 \\ 0 & 0 & 0 \\ 0 & 0 & 1\end{array}\right]$ where $\gamma$ is the reference frictional constant which is set to be unity in the following discussions. Since two frictional constants are equal to each other, two ends of the chain are symmetrically coupled to two baths. We expect that the effective temperature should be different from the mean temperature of two baths since the first symmetry condition mentioned at the end of subsection~\ref{subsec-twooscillators} is broken in this model.

Taking $\mathbf{E}_L=\left[\begin{array}{ccc}1 & 0 & 0\\ 0&0&0\\ 0&0 & 0\end{array}\right]$, we can obtain the diagonal elements of $\mathbf{G}_{L}$:
\begin{equation}G_{L11}=313/385,~G_{L22}={1}/{5},~G_{L33}=72/385.
\end{equation}
Similarly, with the consideration of $\mathbf{E}_R=\left[\begin{array}{ccc}0 & 0 & 0\\ 0&0&0\\ 0&0 & 1\end{array}\right]$, we can derive the diagonal elements of $\mathbf{G}_{R}$:
\begin{equation}
G_{R11}=72/385,~G_{R22}={4}/{5},~G_{R33}=313/385.
\end{equation}

Substituting these diagonal elements into Eqs.~(\ref{eq-weightfactor2}) and (\ref{eq-weightfactor3}), we have
\begin{equation}
C_L=\frac{2 }{5},~\mathrm{and}~C_R=\frac{3}{5}.
\end{equation}
Correspondingly, the effective temperature is 
\begin{equation}T_e=\frac{2 T_L+3 T_R}{5}.\end{equation}
Just as we expect, the effective temperature is indeed different from the mean temperature $\bar{T}\equiv (T_L+T_R)/2$ since the first symmetry condition mentioned at the end of subsection~\ref{subsec-twooscillators} is broken in this model.

With the consideration of Eq.~(\ref{eq-intenergy2}), we derive the internal energy $\mathcal{E}=3T_e=3(2 T_L+3 T_R)/5$. This result can be rewritten as
\begin{equation}\mathcal{E}=3\bar{T}-\frac{3\Delta T}{10},\label{interenegywuwang}\end{equation}
which is identical to Eq.~(95) obtained by Wu and Wang in Ref.~\cite{Wu-Wang2022}. In other words, the internal energy can be expressed in a concise form with the weighted average temperature, while still leads to the correct result.

\end{document}